\documentclass[prd,nofootinbib,preprintnumbers,twocolumn,a4paper]{revtex4}

\pdfoutput=1
\usepackage{latexsym, amssymb, hyperref, graphicx, slashed, color, multirow}
\usepackage[T1]{fontenc}
\definecolor{nicered}{rgb}{0,0.6,0.4}
\definecolor{nicegreen}{rgb}{.1,.5,.1}
\definecolor{darkblue}{rgb}{0,0,.8}
\hypersetup{colorlinks, citecolor=nicered ,linkcolor=darkblue, urlcolor=nicered}

\usepackage{graphicx}
\usepackage{amsmath,amssymb}
\usepackage{amsfonts}
\usepackage{latexsym}
\usepackage{mathrsfs}
\usepackage{adjustbox}
\usepackage{color}
\usepackage{slashed}
\usepackage{float}
\usepackage{feyn}
\usepackage[all]{xy}
\usepackage{feyn}
\usepackage{hyperref}
\usepackage{dsfont}
\usepackage{placeins}
\usepackage{cancel}
\usepackage[top=0.40in, bottom=0.3 in, left=0.8 in, right=0.8 in]{geometry}

\allowdisplaybreaks

\definecolor{blue}{rgb}{0,0,0.5}
\definecolor{lightblue}{rgb}{0,0,1}
\definecolor{red}{rgb}{0.5,0,0}
\definecolor{lightred}{rgb}{1,0.5,0}
\definecolor{green}{rgb}{0,0.5,0}
\definecolor{darkgreen}{rgb}{0.0,0.3,0.0}
\definecolor{orange}{rgb}{1,0.4,0}
\definecolor{grey}{rgb}{0.5,0.5,0.5}
 
 \providecommand{\keywords}[1]
{
  \small	
  \textbf{Keywords :} #1
}

\newcommand{\printifnonempty}[2]{\if\relax\detokenize{#1}\relax\else#2\fi}
\newcommand{\alter}[5]{%
  \long\def\temp{#3}%
  \long\def\accept{#5}%
  \ifx\temp\accept
    {#1}
  \else
    {\textcolor{#4}{\printifnonempty{#1}{{#1}}}%
    \textcolor{grey}{\printifnonempty{#2}{(#2)}}%
    \textcolor{#4}{\printifnonempty{#3}{{[#3]}}}}%
  \fi
}

\usepackage{adjustbox}

\definecolor{MyGrey}{rgb}{0,0,0} 
\definecolor{MyDarkBlue}{rgb}{0.,0.,1} 
\definecolor{MyLightBlue}{rgb}{0.22,0.51,0.9}

\usepackage{mathrsfs,amssymb,slashed}  
\usepackage{cancel}
\usepackage[normalem]{ulem}

\begin{document}
\preprint{\textbf{OSU-HEP-17-05}}

\title{New Physics Scale from Higgs Observables with Effective Dimension-6 Operators}

\author{\textbf{Sudip Jana$^{a,b}$}}
\email{sudip.jana@okstate.edu}
\author{\textbf{S. Nandi$^{a,c}$}}
\email{	s.nandi@okstate.edu, sn35@rice.edu}

\affiliation{$^{a}$ Department of Physics and Oklahoma Center for High Energy Physics,
Oklahoma State University, Stillwater, OK 74078-3072, USA \\ $^{b}$Theory Department, Fermi National Accelerator Laboratory, P.O. Box 500, Batavia, IL 60510, USA \\ $^{c}$ Department of Physics, Rice University, Houston, TX 77034  }

\begin{abstract}

\noindent
No matter what the scale of new physics is,  deviations from the Standard Model (SM) for the Higgs observables will indicate the existence of such a scale. We consider effective six dimensional operators, and their effects on the Higgs productions and decays to estimate this new scale.  We analyze and identify the parameter space consistent with known properties of the Higgs boson using recent Run II results from ATLAS and CMS experiments corresponding to  $\sim$ 37 fb$^{-1}$ of data. We then calculate the  $\bar{t} t h$ productions , as well as double Higgs production at the LHC using the  effective couplings, and show that these can be much different than those predicted by the Standard Model, for a wide region of allowed parameter space. These predictions can be tested in the current or the future runs of the LHC. We find that the data are consistent with the existence of a new physics scale as low as 500 GeV for a significant region of parameter space of this six dimensional couplings with these new physics effects at the LHC. We also find that for some region of the parameter space, di-Higgs production can be much larger than that predicted by the Standard Model, giving rise to the prospect of its observation even in the  current run II of the LHC.
\end{abstract}
\maketitle

\vspace{-0.2 in}
\section{\label{sec:Introduction}Introduction}
\vspace{-0.21 in}

There have been several  major discoveries in the past few decades  culminating with the observation of the Higgs boson in 2012 \cite{Aad:2012tfa,Chatrchyan:2012xdj}. This  is a tremendous success of the SM. However, as most of us agree,  SM can not be the whole story. The Higgs production  in various modes and its decays into various final states so far agrees with the SM. But  uncertainties with the SM predictions still remain in some of the observables of these measurements. This encourages us to venture into the possibility of a new physics  scale that might be estimated from the uncertainty in these measurements.  Also, using this approach,  we might be able to make  predictions which can be tested at the LHC. With this aim in mind,  we consider the effect of a selected set of dimension six operators  relevant for the Higgs Physics, in addition to the contribution from the SM. The dimension six operators related to the Higgs physics  can be introduced both in the strong sector, as well as in the electroweak sector.  Such operators will make extra contributions for the Higgs productions, as well as for its various decay modes. In the most general case, for the effective dimension six operators, there are many operators, and involve large number of parameters. In order to reduce the number of parameters, we only consider a selected set of such operators in the gauge sector (both strong and electroweak (EW)), as well as in the Yukawa sector. In particular, we include only those operators which are  responsible for larger effects, and do not affect the constraints from the EW precision tests in a significant way.

The effective field theory provides a model independent framework for interpreting precision  measurements  connecting  to  specific  UV  models  systematically \cite{new1}. Constraints on these operators have been derived from electroweak (EW) precision measurements \cite{new2,new3,new4}, Higgs sector measurements \cite{new7,new8,new9,new10} and from the triple gauge couplings \cite{new5, new6}.  Using EW data, global fits incorporating various searches have been performed in \cite{new11}. Subsequently fits have been performed including Higgs sector constraints \cite{new12,new13,new14,new15}. In this context, di-Higgs production also has been studied here \cite{hh1,hh2,hh3,feyn}. When the Standard Model is considered as an effective low-energy theory, higher dimensional interaction terms appear in the Lagrangian. Dimension-six terms have been enumerated \cite{d61, d62} and there are 15 + 19 + 25 = 59 independent operators (barring flavour structure and Hermitian conjugations). 
However, many of these operators affect processes that are well measured, e.g. flavor physics or electroweak precision observables set strong constraints on  subsets of those operators. Some of them are also not relevant for the Higgs physics observables, which is the main emphasis of this work. 
Here, we focus on the effective operators that focus on the Higgs physics, and nothing else. This is in the spirit of  reference  \cite{ref2}. 
At the LHC, SM Higgs boson (h) can be produced\footnote{At the 13 TeV LHC, SM Higgs production cross-section via different production modes are summarized as : $\sigma_{ggF}=43.92$ pb, $\sigma_{VBF}=3.748$ pb, $\sigma_{Wh}=1.38$ pb, $\sigma_{Zh}=0.869$ pb, $\sigma_{t\bar{t}h}=508.5$ fb. } significantly via gluon gluon fusion (ggF), vector boson fusion (VBF), associated production with W and Z bosons (Vh) or in association with $t\bar{t}$ ($t\bar{t}h$). Due to insertion of the dimension-6 terms, SM Higgs production as well as decay branching ratios can be largely affected in these production modes.  (1) In the single Higgs production , the most important is the coupling of the gluon pairs to the Higgs boson. Here we have the contribution from the SM dimension-4 operators contributing via the top quark loop. There may exist effective dimension 6 operator (contact interaction) emerging from new physics contributing to this production. (2) The Yukawa coupling of the top quark to the Higgs boson is most important in single Higgs production. Here also, there may exist dimension-6 operator (in addition to the dimension 4 present in the SM) emerging again from the new physics. This will also affect the $t \bar{t} h$ production, as well as the double Higgs productions, which are of great importance in the upcoming LHC runs.  (3) In the production of the Higgs boson in association with W or Z,  the important contribution of dimension-6 operator will be the $hZZ$ or $hWW$ couplings, which will further effect the decays of the Higgs to $W W^*$ and $Z Z^*$. Thus, in addition to the contribution from the usual SM, the contribution of the effective dimension six operators will be important here.  (4) The dominant decay mode of the Higgs boson is to $b \bar{b}$, the branching ratio being  $ \simeq 60\%$. Thus the dimension-6 contribution to the Yukawa coupling of the Higgs to the bottom pairs will also be very important to look for a new physics scale in the Higgs observables. (5) We have also included dimension-6 operator in the Higgs potential. This has the largest effect on our results on the di-Higgs productions, since it changes the effective triple Higgs coupling in a major way. Using the above five criteria, we narrow down our analysis to include five new parameters and these are $g^{(6)}$, $y_{t}^{(6)}$ , $y_{b}^{(6)}$, $y_{g}^{(6)}$, $\lambda^{(6)}$ and the new physics scale , M.  We have done the analysis also including dimension-6 tau Yukawa term $y_{\tau}^{(6)}$. As the branching ratio of the Higgs in the $ \tau \tau$ mode is $\simeq 6\%$, it does not significantly affect the major Higgs observables and hence, the phenomenology we are concentrating. We ignore its contribution for rest of our analysis. A complete list of all effective dimension-6 operators can be found in \cite{d61, d62}. 

With these above assumptions, we first identify the parameter space consistent with the Higgs observables and then we find two important results. (1) The   $t \bar{t} h$ coupling  can be much larger or smaller  than that predicted by the SM, and thus  giving rise to significantly different rate of  $t \bar{t} h$  productions. (2) Double Higgs productions can be much larger than that predicted by the SM.


Very recently, the CMS collaboration has reported a search for the production of a Standard Model (SM) Higgs boson in association with a top quark pair ($t\bar{t}h$) at the LHC Run-2 and a best fit $t\bar{t}h$ yield of $1.5\pm 0.5$ times the SM prediction with an observed significance of $3.3\sigma$ \cite{tthcms}, whereas ATLAS reported limit is $1.8\pm 0.7$ \cite{tthatlas} on $t\bar{t}h$ production. ATLAS and CMS reported signal strength values are consistent with the SM. However, the central values of the signal strength $\mu_{t\bar{t}h}$ is significantly different from one. There are several literatures \cite{tth1}  attempting to explain the issue for the enhanced $t\bar{t}h$  production. As we shall see, in our framework, the signal strength $\mu_{t\bar{t}h}$ can be as large as 2.4 and also as low as 0.5. There are still large uncertainties in the $t\bar{t}h$ measurements. If any significant deviation (enhancement or suppression) arises in $t\bar{t}h$ production rate at the LHC, this is the best model independent approach to explain the scenario. On the other hand, ATLAS and CMS collaborations have reported the new results on di-Higgs boson searches \cite{econf, mor,CMShh,CMShh2,ATLAShh} looking at the different final states ($b\bar{b}\gamma\gamma, b\bar{b}\tau^{+} \tau^{-}, b\bar{b}b\bar{b}$ and  $b\bar{b}W^+W^-$), using 36 fb$^{-1}$ data from Run II of LHC at 13 TeV. No signal has been observed and the stringent limit of 646 fb on the di-Higgs production cross section is reported \cite{econf, mor,CMShh,CMShh2,ATLAShh}. In SM, $hh$ production cross-section is  about $33.45$ fb. After considering effective dimension six couplings, according to our analysis, the di-Higgs production can be as large as about $\sim 636 $ fb, which is $19$ times of the SM predicted cross-section for some region of the six dimensional parameter space. If nature does realize this parameter space, di-Higgs production may be observable even at the current run 2 of the LHC as more data are accumulated.

The paper is organized as follows: In Sec. \ref{sec:2}, we discuss the formalism and analyze the dimension-6 operators. Thereafter in Sec. \ref{collider}, we perform the numerical simulations for collider signatures. Finally we conclude.  

\vspace{-0.65cm}
\section{Formalism} \label{sec:2}
Our gauge symmetry is the same as the SM. We are introducing a selected set  of additional dimension six operators which can affect the Higgs observables in a major way. These operators are all invariant under the SM gauge symmetry. 
\vspace{0.1cm}

\textbullet { \textbf{EW Yukawa sector}}:  
\begin{align} 
\mathcal{L}_{Yuk}^{(6)} \supset \frac{y_{t}^{(6)}}{M^2} (\bar{t}_{L}, \bar{b}_{L}) t_R \tilde{H} (H ^{\dagger} H) +\frac{y_{b}^{(6)}}{M^2} (\bar{t}_{L}, \bar{b}_{L}) b_R H (H ^{\dagger} H)  \nonumber \\ +\frac{y_{\tau}^{(6)}}{M^2} (\bar{\nu}_{\tau}, \bar{\tau}_{L}) \tau_R H (H ^{\dagger} H) + h.c.
\end{align}
 We have included the dimension-6 terms for third generation fermions only. For simplicity, we have included only the flavor diagonal dimension six Yukawa couplings. Similarly, we can extend it for first and second generation fermions also. But, since we are interested in new physics affecting Higgs rates in a major way, we ignore the negligible effects originating from dimension-6 Yukawa terms for first and second generation fermions. We will also ignore the dimension six operator for the $\tau$ lepton. The Higgs branching ratio to $ \tau$ pair is very small $6\%$, and its inclusion does not affect the phenomenology we are concentrating.
   
\textbullet  { \textbf {Strong sector}}:   
\begin{equation}
  \mathcal{L}_{Strong}^{(6)} \supset \frac{g^{(6)}}{M^2}G^{\mu \nu a}  G_{\mu \nu a} ( H^{\dagger} H )
\end{equation} 

This operator will contribute to the Higgs production , as well as its decay to two gluons.  $g^{(6)}$ is an unknown parameter, and M is the new physics scale. This operator (the contact term)  will significantly  contribute, in addition to the SM contribution via the top quark loop, in single Higgs production via gluon gluon fusion process.

\textbullet  { \textbf {EW gauge sector}}: 
\begin{equation}\label{eq:ewgauge}
 \mathcal{L}_{EW gauge}^{(6)} \supset  \frac{y_{g}^{(6)}}{M^2}(D^{\mu} H)^\dagger  (D_{\mu}  H)   (H^{\dagger} H)
\end{equation} 
where  the coupling $ y_{g}^{(6)}$ is an arbitrary coefficient \footnote{ For simplicity we focus on CP-conserving operators, CP-violating ones can be included in a straightforward way. We omit the operator $\mid H^{\dagger}  D^{\mu} H \mid ^2$, since it violates the custodial symmetry and is strongly constrained by LEP data. Its inclusion has no impact on our analysis.}. There are several other dimension six operators which we neglect. The reason is that they do not contribute in a significant way to the processes we are emphasizing on this work, and  some of them, if the coefficients are not very small, may mess up the EW precision test. We discuss briefly the effect of this operator above for the processes of interest. This operator contributes to the decays of $h \to W W^{\star}, Z Z^{\star}$ as well as to the production through VBF and associated Higgs production with W or Z boson.

\textbullet  { \textbf {Scalar Potential}}:   
\begin{equation}
  \mathcal{L}_{Scalar}^{(6)} \supset \frac{\lambda^{(6)}}{M^2} ( H^{\dagger} H )^3
\end{equation} 

This operator will modify the Higgs trilinear coupling, and hence, contribute  significantly to the di-Higgs production. 

Note that in Eq.\ref{eq:ewgauge}, when we put the VEV of the Higgs boson
, this operator modifies \cite{kin} the Higgs kinetic term  $\frac{1}{2}\partial^\mu h \partial_\mu h$ to $\left(1+\frac{y_{g}^{(6)}v^2}{2M^2}\right)\frac{1}{2}\partial^\mu h \partial_\mu h$. (Throughout our analysis, we use the convention $H=\begin{pmatrix} 0 \\ \frac{h+v}{\sqrt{2}} \end{pmatrix}$ in unitary gauge). Hence, we need to redefine the Higgs field by dividing out with the factor $N=\left(1+\frac{y_{g}^{(6)}v^2}{2M^2}\right)^{1/2}$ to get the canonically normalized form for the kinetic term $\frac{1}{2}\partial^\mu h \partial_\mu h$. This modifies the usual couplings of the Higgs field to the gauge bosons , the fermions  and the  Higgs bosons as given below.

\begin{eqnarray}
\kappa_{V} &=& \left[ \frac{1}{N^2}+ \frac{y_{g}^{(6)}v^2}{M^2 N^4}\right], \\
\kappa_t &=& \left[\frac{1}{N} +\frac{y_{t}^{(6)}v^3}{\sqrt{2}m_t M^2 N^3}  \right], \\
\kappa_b &=& \left[\frac{1}{N} +\frac{y_{b}^{(6)}v^3}{\sqrt{2}m_b M^2 N^3}  \right], \\
\kappa_\tau &=& \left[\frac{1}{N} +\frac{y_{\tau}^{(6)}v^3}{\sqrt{2}m_{\tau} M^2 N^3}  \right], \\
\kappa_g &=& \frac{\left[1.034\kappa_t + \epsilon_b \kappa_b + \frac{4 \pi g^{(6)} v^2}{\alpha_s N^2 M^2} \right]}{\left[1.034 + \epsilon_b \right]} , \\
\kappa_{hhh} &=& \left[ \frac{1}{N^4}- \frac{5\lambda^{(6)}v^4}{m_h^2 M^2 N^6}\right], \\
\kappa_{\gamma\gamma} &=&  \left| \frac{\frac{4}{3} \kappa_t F_{1/2}(m_h) + \kappa_V F_1(m_h) }{\frac{4}{3} F_{1/2}(m_h) + F_1(m_h)}  \right|, \\
\kappa_{Z\gamma} &=&  \left| \frac{ \frac{2}{\cos \theta_W} \left( 1- \frac{8}{3} \sin^2 \theta_W \right) \kappa_t F_{1/2}(m_h) + \kappa_V F_1(m_h)}{ \frac{2}{\cos \theta_W} \left( 1- \frac{8}{3} \sin^2 \theta_W \right) F_{1/2}(m_h) + F_1(m_h)}  \right|. \nonumber \\ 
\end{eqnarray}
Loop functions used in this paper are defined as follows:
 \begin{eqnarray}
 \label{eq:loop1}
F_1(x) = -x^2\left[2x^{-2}+3x^{-1}+3(2x^{-1}-1)f(x^{-1})\right]\ ,\\
\label{eq:loop2}
F_{1/2}(x) = 2  \, x^2 \left[x^{-1}+ (x^{-1}-1)f(x^{-1})\right] \ , \\
\epsilon_b = -0.032+0.035 i\,
\end{eqnarray}
For a Higgs mass below the kinematic threshold of the loop particle, $m_h < 2 \; m_{\rm loop}$, we have
 \begin{equation}
 f(x) = \arcsin^2 \sqrt{x}  \, 
 \end{equation}
 where $x_i \equiv 4m_i^2/m_h^2$ ($i=t, W$).

We now calculate the partial decay widths for various SM Higgs decay modes : 
\begin{gather}
\Gamma_{h \to \gamma\gamma} = \kappa_{\gamma\gamma}^2 \Gamma_{h \to \gamma\gamma}^{{\rm SM}}, \\
\Gamma_{h \to WW^*} = \kappa_V^2 \Gamma_{h \to WW^*}^{{\rm SM}}, \\
\Gamma_{h \to ZZ^*} = \kappa_V^2 \Gamma_{h \to ZZ^*}^{{\rm SM}}, \\
\Gamma_{h \to b\bar{b}} = \kappa_b^2 \Gamma_{h \to bb}^{{\rm SM}}, \\
\Gamma_{h \to \tau^+\tau^-} = \kappa_\tau^2\Gamma_{h \to \tau\tau}^{{\rm SM}}, \\
\Gamma_{h \to gg} = \kappa_g^2 \Gamma_{h \to gg}^{{\rm SM}}, \\
\Gamma_{h \to Z\gamma} = \kappa_{Z\gamma}^2 \Gamma_{h \to Z\gamma}^{{\rm SM}},
 \end{gather}
where the partial decay widths in the SM can be found in \cite{Gunion}. 
\section{Collider Phenomenology}\label{collider}
\vspace{-0.2cm}
In this section, we study the collider phenomenology of the Higgs sector. 
In particular, we  discuss the possibility if the effective dimension-6 operators within this framework can explain the significant deviation in  $t \bar{t} h$ production cross section, as recently indicated by CMS \cite{tthcms} and ATLAS collaboration \cite{tthatlas},  along with the other Higgs boson properties.   We also  want to investigate if the di-Higgs production may be observable at the current or future runs of the LHC. 

 To start this effort, we first numerically analyze the effects of dimension-6 terms on the $t \bar{t} h$ production as well as the signal strengths of Higgs boson decay modes for  $h \to \gamma\gamma, WW, ZZ, b\bar{b}, \tau \bar{\tau}, Z\gamma$.  Then, we identify a parameter space which is consistent with both the recent ATLAS and CMS results on the LHC Run-1 and Run-2 ($37$ $fb^{-1}$) data.  Then remaining within the allowed parameter space, we analyze the possible  signals,  such as the  enhanced di-Higgs boson production that may be observable at the current or future run of the LHC. The relevant parameter space of this model is spanned by the three new dimension-6 Yukawa terms, dimension-6 term from electroweak gauge sector, dimension-6 term from strong sector, dimension-6 term from scalar potential and the mass of the new physics scale :
\begin{equation}
    \left\lbrace y_{t}^{(6)},\hspace*{0.2cm} y_{b}^{(6)},\hspace*{0.2cm} y_{g}^{(6)}, \hspace*{0.2cm} g^{(6)},  \hspace*{0.2cm} \lambda^{(6)}, \hspace*{0.2cm} M \right\rbrace
\label{eq:parameters}
\end{equation}

In the LHC Higgs observable analysis
\footnote{In our analysis, we employ the center value of the Higgs boson mass $m_h=125.09$ GeV \cite{higgs} and the center value of the combination of Tevatron and LHC measurements of the top quark mass $m_t=173.34$ in GeV \cite{top}.} \cite{higgs}, the searches for Higgs boson at ATLAS and CMS can give strong bounds on these free parameters. The signal strength $\mu$, defined as the ratio of the measured Higgs boson rate to its SM prediction, is used to characterize the Higgs boson yields and it is given by :
\begin{equation}
\mu^i_f =  \frac{\sigma^i \cdot BR_f}{(\sigma^i)_{SM} \cdot (BR_{f})_{SM}} = \mu^i\cdot\mu_f.
\label{eq:muif}
\end{equation}

Here $\sigma^i$  $(i= ggF, VBF, Wh, Zh, t\bar{t}h)$ and $BR_f$  $(f = ZZ^{\star}, WW^{\star}, \gamma \gamma, \tau^+ \tau^-, b\bar{b}, \mu^+ \mu^-)$ are respectively the SM Higgs production cross section for different production mechanism ($i\to h$) and the branching fraction for different decay modes of SM Higgs ($h\to f$).

\begin{table}[]
\renewcommand{\arraystretch}{1.5}
\begin{center}
\adjustbox{max height=\dimexpr\textheight-9.0cm\relax,
           max width=\dimexpr\textwidth-3.0cm\relax}{
\begin{tabular}{| c || c | c | c | }
\hline \hline
\centering 
\multirow{2}{*}{\shortstack{Decay \\ channel}}  & \multirow{2}{*}{\shortstack{Production \\ Mode}} & \multirow{2}{*}{CMS}  & \multirow{2}{*}{ATLAS}  \\ 
&  & & \\ \hline \hline
\multirow{4}{*}{$\gamma \gamma$} & $ ggF $ & $\rm 1.05^{+0.19}_{-0.19}$ \cite{gammarun21}    & $\rm 0.80^{+0.19}_{-0.18}$ \cite{gammarun22} \\
& $VBF$ & $\rm 0.6^{+0.6}_{-0.5}$ \cite{gammarun21}  & $\rm 2.1^{+0.6}_{-0.6}$ \cite{gammarun22} \\
& $Wh$ & $\rm 3.1^{+1.50}_{-1.30}$ \cite{gammarun21} & $\rm 0.7^{+0.9}_{-0.8} $\cite{gammarun22} \\
& $Zh$ & $\rm 0.0^{+0.9}_{-0.0} $\cite{gammarun21} & $\rm 0.7^{+0.9}_{-0.8}$ \cite{gammarun22} \\
\hline
\multirow{4}{*}{$ZZ^{\star}$} & $ ggF $ & $\rm 1.20^{+0.22}_{-0.21} $\cite{zrun21}    & $\rm 1.11^{+0.23}_{-0.27} $\cite{zrun22} \\
& $VBF$ & $\rm 0.05^{+1.03}_{-0.05}$ \cite{zrun21}  & $\rm 4.0^{+2.1}_{-1.8} $\cite{zrun22} \\
& $Wh$ & $\rm 0.0^{+2.66}_{-0.00}$ \cite{zrun21} & $\rm <3.8$\cite{zrun22} \\
& $Zh$ & $\rm 0.0^{+2.66}_{-0.00}$ \cite{zrun21} & $\rm <3.8$ \cite{zrun22}
\\ \hline
\multirow{4}{*}{$W^{+}W^{-}$} & $ ggF $ & $\rm 0.9^{+0.40}_{-0.30} $\cite{w3}\textsuperscript{\ref{fn1}}   & $\rm 1.02^{+0.29}_{-0.26} $  \cite{w1}\footnote{{\label{fn1}}Results from 36 fb$^{-1}$ data from 13 TeV LHC is not still reported.}  \\
& $VBF$ & $\rm 1.4^{+0.8}_{-0.8} $\cite{w3}\textsuperscript{\ref{fn1}}  & $\rm 1.7^{+1.1}_{-0.9} $ \cite{w2}\textsuperscript{\ref{fn1}}  \\
& $Vh$ & $\rm 2.1^{+2.3}_{-2.2} $\cite{w3}\textsuperscript{\ref{fn1}}  & $\rm 3.2^{+4.4}_{-4.2} $ \cite{w2}\textsuperscript{\ref{fn1}}  \\
& $ ggF+VBF+Vh $ & $\rm 1.05^{+0.26}_{-0.26} $\cite{w3}\textsuperscript{\ref{fn1}}   & -  \\ \hline 
\multirow{1}{*}{$b\bar{b}$} & $Vh$ & $\rm 1.06^{+0.31}_{-0.29} $\cite{b2}  & $\rm 0.9^{+0.28}_{-0.26} $ \cite{b1}  \\ \hline
\multirow{3}{*}{$\tau^{+} \tau^{-}$} & $ggF$ & $\rm 1.05^{+0.49}_{-0.46}$ \cite{tau2}  & $\rm 2.0^{+0.8}_{-0.8} $ \cite{tau1}\textsuperscript{\ref{fn1}}  \\
& $VBF+Vh$ & $\rm 1.07^{+0.45}_{-0.43}$ \cite{tau2}  & $\rm 1.24^{+0.58}_{-0.54}$ \cite{tau1}\textsuperscript{\ref{fn1}}  \\
& $ggF+VBF+Vh$ & $\rm 1.06^{+0.25}_{-0.24}$ \cite{tau2}  & $\rm 1.43^{+0.43}_{-0.37}$ \cite{tau1}\textsuperscript{\ref{fn1}}  \\ \hline \hline 
\end{tabular}
}
\caption{Signal strength constraints from recently reported 13 TeV 36 $fb^{-1}$ LHC data along with references. }
\label{table-mu8}
\end{center}
\end{table}

\begin{figure}
$$
 \includegraphics[height=4.55cm, width=0.25\textwidth]{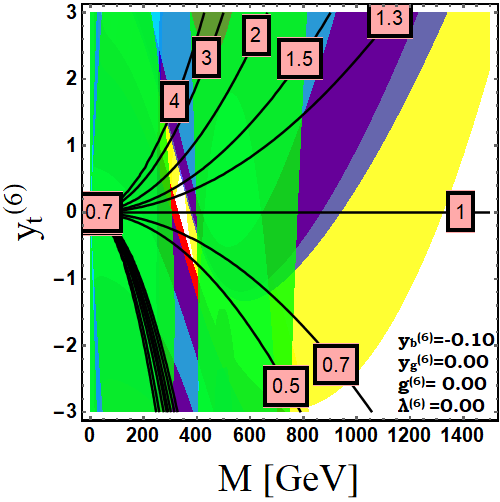}
  \includegraphics[height=4.55cm, width=0.25\textwidth]{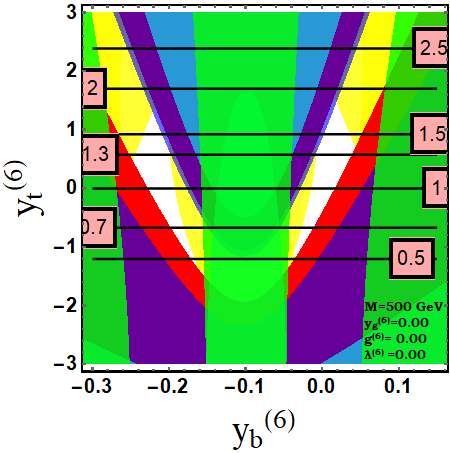}
 $$
 $$
 \includegraphics[height=4.55cm, width=0.25\textwidth]{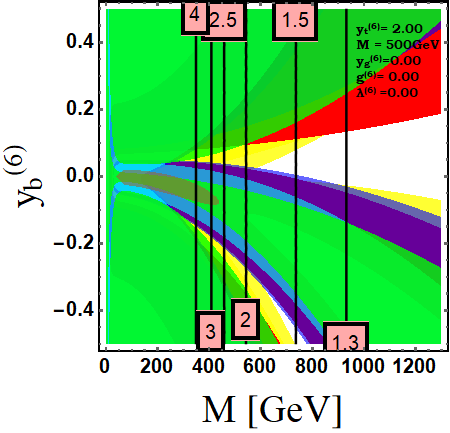}
 $$
 \caption{Top Left : Contour plot of $\mu^{t\bar{t}h}$ in $\lbrace y_{t}^{(6)}, M\rbrace $ plane; Top Right : Contour plot of $\mu^{t\bar{t}h}$ in $\lbrace y_{t}^{(6)}, y_{b}^{(6)}\rbrace $ plane and Bottom : Contour plot of $\mu^{t\bar{t}h}$ in $\lbrace{y_{b}^{(6)}}, M\rbrace $ plane. The yellow, cyan, green, red and purple shaded regions are excluded from the signal strength limits [cf. Table \ref{table-mu8}] for various decay modes ($\gamma\gamma, \tau \tau, b\bar{b}, ZZ^{\star}, WW^{\star}$) respectively at 95$\%$ confidence level. The white shaded region simultaneously satisfies all the experimental constraints. Boxed numbers indicate the $\mu_{t\bar{t}h}$  values.  
 }
 \label{fig1}
\end{figure}

The ATLAS and CMS run 1 data are combined and analyzed using the signal strength formalism and the results are presented in \cite{higgs}. Recently, ATLAS and CMS collaborations have updated the results \cite{econf} on Higgs searches based on 37 $fb^{-1}$ data at 13 TeV LHC. The individual analyses examine a specific Higgs boson decay mode, with categories related to the various production processes and they are $h \to \gamma\gamma$ \cite{gammarun21,gammarun22,gamma1,gamma2}, $h \to ZZ^\star$ \cite{zrun21,zrun22,z1,z2}, $h \to WW^\star$ \cite{w1,w2,w3}, $h \to \tau \tau$ \cite{tau1,tau2}, $h \to b\bar{b}$ \cite{b1,b2} and $h \to Z\gamma$ \cite{zgamma1,zgamma2}. Throughout our study, we have used the most updated ATLAS and CMS reported results on 125 GeV Higgs boson searches to impose constraints on signal strengths for various decay modes at 95$\%$ confidence level and which is summarized in Table \ref{table-mu8}.

For our analysis, we adopt the following strategy. 

 (1) First, we introduce dimension-6 operator in the Yukawa sector and try to explore whether any new physics effect (enhanced/suppressed couplings of Higgs to fermions) can be achieved satisfying all Higgs physics constraints and try to identify the six dimensional parameter space where these effects can arise.
 
  (2) Then we introduce dimension-6 operator in the EW gauge sector and discuss its effect following the previous effects from the Yukawa sector.
  
   (3) After that we introduce dimension-6 term in strong sector and analyze both individual and combined effects of all of these dimension-6 operators and discuss the new physics effects.
   
    (4) Then, we introduce dimension-6 operator in the scalar potential and analyze its effect in di-Higgs production. 
    
    (5)Finally, we discuss about two correlated new physics signatures :  enhanced  (or suppressed) $t\bar{t}h$ and enhanced $hh$ production.

 Since the gauge structure of the SM has been very well established from the precision measurements, as mentioned above, we  first concentrate on the Yukawa sector, in particular, the effects coming from the six dimensional Yukawa couplings for the third generation fermions. 
 The top and bottom Yukawas ($y_{t}^{(6)}$ and $y_{b}^{(6)}$) play key roles in Higgs observable. The top Yukawa dictates the production of SM Higgs mostly, whereas the bottom Yukawa guides the branching ratio for different decay modes of SM Higgs h.  Since the partial decay width for $h \to b\bar{b}$ mostly contributes $\sim 58 \%$ to the total Higgs decay width, any slight deviation in bottom Yukawa will change the total decay width and hence the branching ratio to other decay modes. We analyze the full parameter space of extra Yukawa terms and new physics scale affecting the SM Higgs physics and impose constraints from the signal strength limits [cf. Table \ref{table-mu8}] for various decay modes ($\gamma\gamma, \tau \tau, b\bar{b}, ZZ^{\star}, WW^{\star}$) at 95$\%$ confidence level.  The effect is displayed in Fig.1. The white shaded region simultaneously satisfies all the experimental constraints. Since $y_{\tau}^{(6)}$ has no significant contribution to the total decay width of SM Higgs compared to $y_{b}^{(6)}$,  as mentioned before, we have ignored $y_{\tau}^{(6)}$ for our analysis regarding the effect of dimension six operators. It does not affect the phenomenology we are concentrating. 
 
 Next, we evaluate the signal strength $\mu^{t\bar{t}h}$ $(= \kappa_t^2)$ for the production of SM Higgs associated with the top quark pair. Upper left segment of Fig. \ref{fig1} shows the contour plot of $\mu^{t\bar{t}h}$ in $\lbrace y_{t}^{(6)}, M \rbrace$ plane for a fixed value of $y_{b}^{(6)}(=-0.1)$, whereas upper right segment shows the contour plot of $\mu^{t\bar{t}H}$ in $\lbrace y_{t}^{(6)}, y_{b}^{(6)}\rbrace$ plane for a fixed value of $M = 500$ GeV and bottom one of Fig. \ref{fig1} shows the contour plot of $\mu^{t\bar{t}h}$ in $\lbrace y_{b}^{(6)}, M \rbrace$ plane for a fixed value of $y_{t}^{(6)}(=2)$.   Fig. \ref{fig1} clearly indicates that within this framework, $t\bar{t}h$ can be produced up to 2 times of the SM  predicted cross-section at the LHC satisfying all the current experimental constraints from 125 GeV Higgs boson searches while we allow a variation of $y_{t}^{(6)}$ between -3 to 3.  On the other hand, $t\bar{t}h$ production rate can also be as low as 0.5 times weaker than the SM predicted value. This enhanced or suppressed $t\bar{t}h$ production can be the new physics signature and it can be tested at the LHC. We mention that, although SM Higgs h is resonantly produced in gluon gluon fusion via triangular loop circulated by top quarks mainly, there is small effect ($\sim 7 \%$) due to the bottom quark circulated loop. When bottom Yukawa comes up with negative sign, its effect becomes larger ($15\%$) and we consider that effect too. Due to the different interference pattern between Yukawas ($y_{t}^{(6)}, y_{b}^{(6)}$) in production as well as in decay modes, these plots are not symmetric about the central axes.  We have also calculated the signal strength for $ Z \gamma$ channel and  which is consistent with the available experimental data \cite{zgamma1,zgamma2}. The signal strength in $ Z \gamma$ channel can be achieved from 0.6 to 1.5 satisfying all the constraints. There are models beyond the SM which predict this type of anomalous Yukawa couplings of the physical Higgs boson, such as the two Higgs doublet model \cite{2hdm,me}. Recently, enhanced $t\bar{t}h$ production and flavor constraints are extensively studied in most general 2HDM \cite{me}. Although we analyze in an effective operator approach, the effect of anomalous Yukawa couplings due to dimension-6 terms is reflected in anomalous Yukawa couplings of SM Higgs due to mixing between two Higgs in 2HDM \cite{me}.  

\begin{figure}[]
$$
 \includegraphics[height=4.5cm, width=0.25\textwidth]{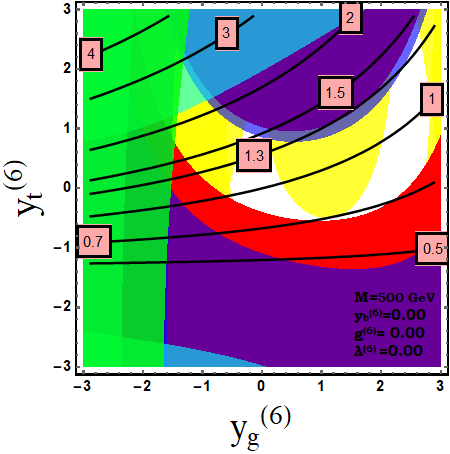}
  \includegraphics[height=4.5cm, width=0.25\textwidth]{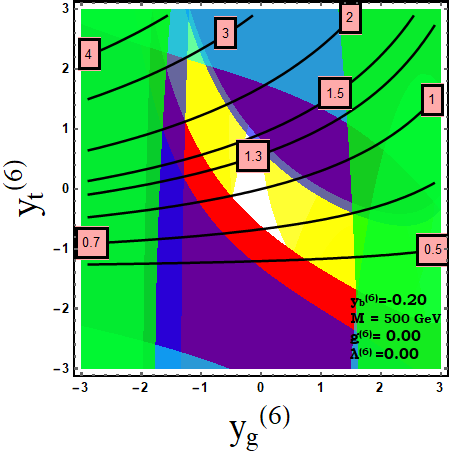}
 $$
 $$
 \includegraphics[height=4.5cm, width=0.25\textwidth]{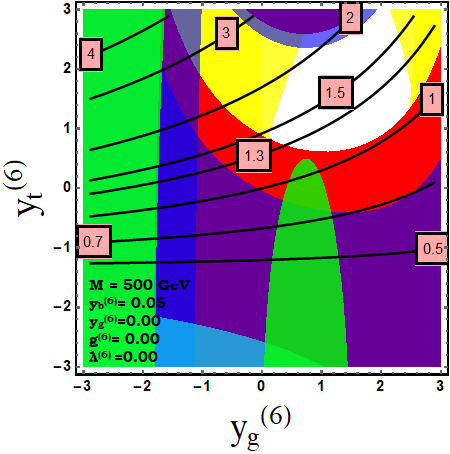}
  $$
 \caption{Contour plot of $\mu^{t\bar{t}h}$ in $\lbrace y_{t}^{(6)}, y_{g}^{(6)}\rbrace $ plane. $y_{b}^{(6)} =$ 0 (upper left) , -0.2 (upper right) and 0.05 (bottom) and the mass scale M is kept fixed at 500 GeV. The yellow, cyan, green, red and purple shaded regions are excluded from the signal strength limits [cf. Table \ref{table-mu8}] for various decay modes ($\gamma\gamma, \tau \tau, b\bar{b}, ZZ^{\star}, WW^{\star}$) respectively at 95$\%$ confidence level. The white shaded region simultaneously satisfies all the experimental constraints. Boxed numbers indicate the $\mu_{t\bar{t}h}$  values.  
 }
\label{2}
\end{figure}

Now, we introduce dimension 6 operator in EW gauge sector and  analyze its effect on Higgs observable. We found that the dimension-6 term, $y_{g}^{(6)}$ in EW gauge sector, is less influential than the dimension-6 terms in Yukawa and strong sectors. In SM Higgs production via ggF process, $y_{g}^{(6)}$ plays no role, whereas branching ratio for $h\to WW, ZZ$ can be modified due to inclusion of $y_{g}^{(6)}$. Fig. \ref{2} depicts the constraints from the signal strength limits [cf. Table \ref{table-mu8}] for various decay modes ($\gamma\gamma, \tau \tau, b\bar{b}, ZZ^{\star}, WW^{\star}$) at 95$\%$ confidence level in $\lbrace y_{t}^{(6)}, y_{g}^{(6)}\rbrace $ plane. We choose $y_{b}^{(6)} =$ 0 (upper left) , -0.2 (upper right) and 0.05 (bottom) and the mass scale M is kept fixed at 500 GeV. As expected and as can be seen from Fig. \ref{2} that as  bottom Yukawa $y_{b}^{(6)}$ gets larger value to enhance overall $b\bar{b}h$ coupling,  $y_{g}^{(6)} $  has to have larger value to satisfy the constraints from Higgs observables. This is due to the fact that, whenever  $y_{b}^{(6)}$ is large, the partial decay width for $h \to b\bar{b}$ mode gets enhanced and hence, total decay width becomes larger suppressing branching ratio for $h \to WW, ZZ$ decay modes. Since  $y_{g}^{(6)}$ has no impact on production via ggF process,  $y_{g}^{(6)}$ has to be larger to enhance the partial decay width for $h \to WW, ZZ$ decay modes making branching ratio almost unaffected to satisfy the correct signal strength limits on ZZ, WW channels. From upper left segment of Fig. \ref{2}, we can see that if dimension-6 terms in Yukawa sector are not introduced and only the effect of $y_{g}^{(6)}$ is considered, we can still get enhanced $t\bar{t}h$ production rate which is almost 1.3 times of the SM predicted value. After inclusion of $y_{t}^{(6)}$ and $y_{b}^{(6)}$, this effect can be much larger and the signal strength for $t\bar{t}h$ production can become as large as 2.4 and as low as 0.5. It is important to mention that whenever $t\bar{t}h$ production is getting enhanced making single Higgs production rate via ggF process larger, overall branching ratios for $h \to WW^{\star}$ or $h \to ZZ^{\star}$ modes has to be suppressed to satisfy correct limits. This also indirectly suppresses the Higgs production in VBF, Wh and Zh processes.  Our scenario predicts enhanced $t\bar{t}h$ production and simultaneously suppressed production of SM Higgs boson in VBF, Wh or Zh processes and this can be tested in the upcoming runs of the LHC. However, there are still large uncertainties in these channels [cf. Table \ref{table-mu8}], but CMS reported central values [cf. Table \ref{table-mu8}] mostly favor this scenario according to the updated status.    

\begin{figure}[]
 $$
 \includegraphics[height=3.5cm, width=0.45\textwidth]{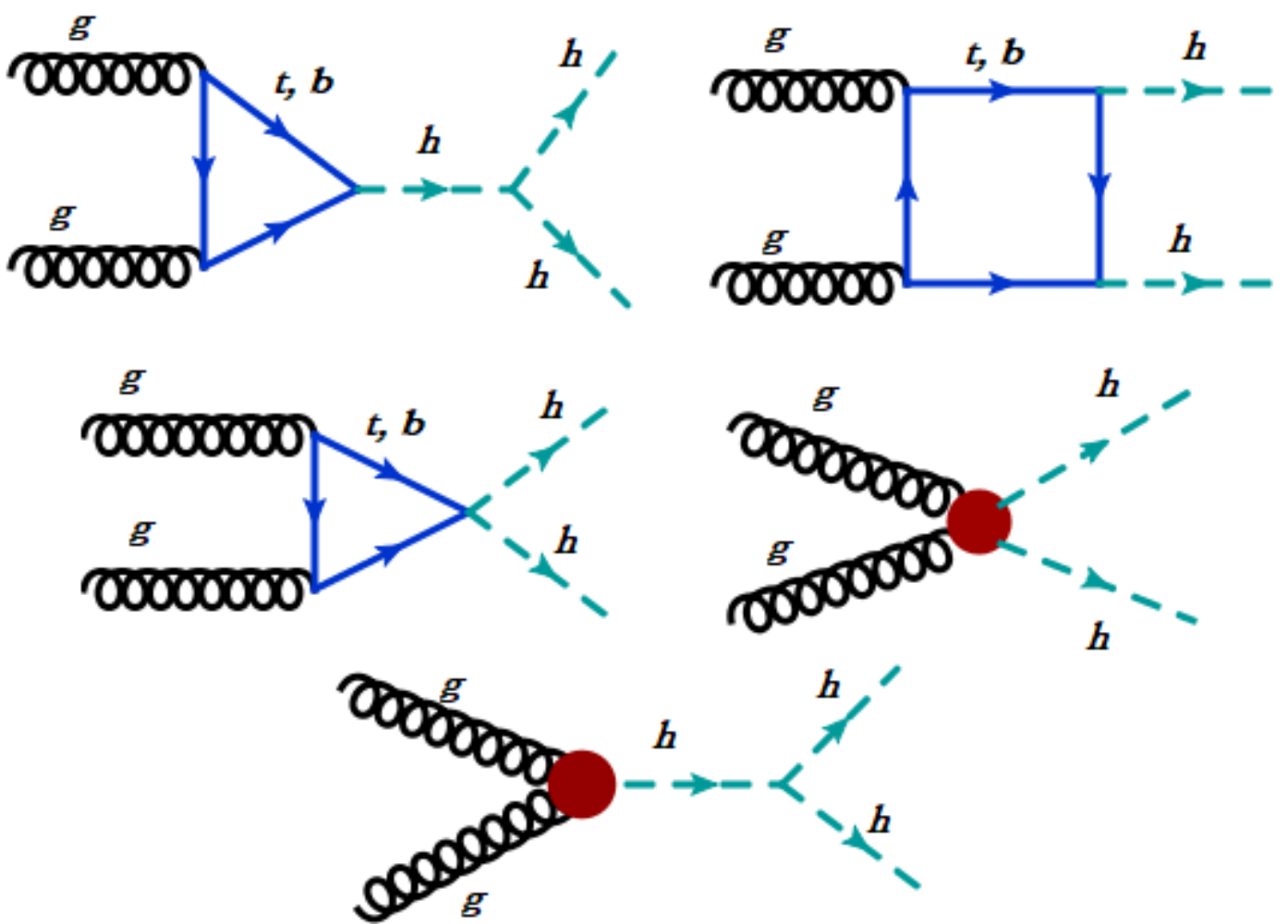}
 $$
 \caption{Feynman diagrams \cite{feyn} contributing to double Higgs production via gluon fusion. The  first two diagrams are present in the SM, while the next three  arise due to dimension 6 operators.
 }
\label{3}
\end{figure}
Next, we introduce the dimension-6 term ($g^{(6)}$) in the strong sector and investigate its effect.  Deviation in di-Higgs production compared to the SM can be one of the new physics effect due to this term. The di-Higgs boson production has drawn a lot of attentions since it is the golden channel to test the EW symmetry breaking mechanism. Since the SM Higgs boson (h) does not carry any color, they are produced in pair through the triangle loop and box loop in SM. The di-Higgs production rate in the SM is small mainly due to the large destructive interference between the triangle and box loop diagrams. At the LHC with a center of mass energy of $13{~\rm TeV}$, the production cross section is about $33.45~{\rm fb}$, which can not be measured owing to the small branching ratio of the Higgs boson decay and large SM backgrounds. The detailed study of SM di-Higgs production can be found in ref.\cite{higgswg}. However, in new physics models, the di-Higgs production cross-section can significantly deviate from the SM value. Due to insertion of the dimension-6 term in strong sector, there will be additional diagrams contributing to the di-Higgs production in addition to the SM contribution and as shown in Fig. \ref{3}. Also, change in SM $t\bar{t}h$ and $hhh$ couplings could give a significant deviation on di-Higgs production cross-section. These two effects could enhance the di-Higgs production and make it testable at the LHC. Therefore, it is important to study how large can the cross section of the double Higgs boson production be considering all the constraints from the single Higgs boson measurements. The $b\bar{b}\gamma\gamma$ final state is particularly promising for this search, as it benefits from the clean diphoton signal due to high $m_{\gamma\gamma}$ resolution and the large branching fraction of the $h\rightarrow b\bar{b}$ decay ($\sim$ 58$\%$). 
We consider the signal strength relative to the SM expectation $\mu_{hh}$  as $\mu_{hh}=\frac{\sigma(pp\to hh)_{New Physics}}{\sigma(pp\to hh)_{SM}}.$

\begin{figure}[]
$$
 \includegraphics[height=4.5cm, width=0.25\textwidth]{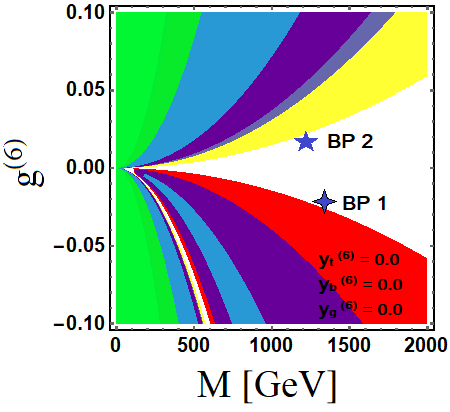}
  \includegraphics[height=4.5cm, width=0.25\textwidth]{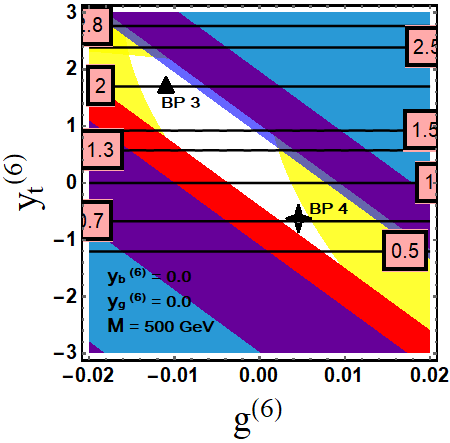}
 $$
 $$
 \includegraphics[height=4.5cm, width=0.25\textwidth]{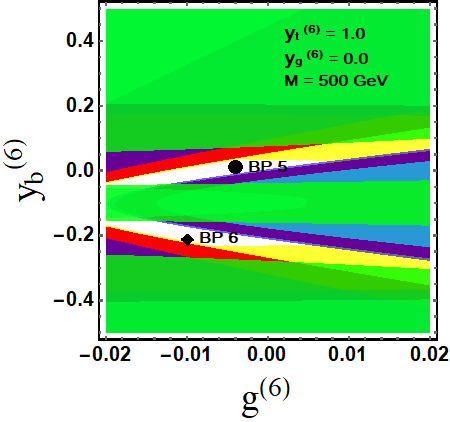}
  \includegraphics[height=4.5cm, width=0.25\textwidth]{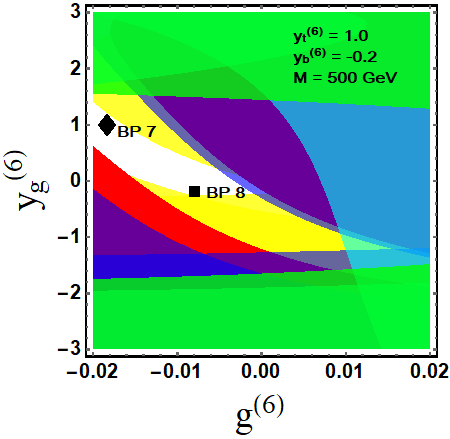}
 $$
 \caption{Constraints in $\lbrace g^{(6)}, M\rbrace $ plane (top left),  $\lbrace y_{t}^{(6)}, g^{(6)}\rbrace $ plane (top right), $\lbrace{y_{b}^{(6)}}, g^{(6)}\rbrace $ plane (bottom left) and  $\lbrace{y_{g}^{(6)}}, g^{(6)}\rbrace $ plane (bottom right) from the signal strength limits [cf. Table \ref{table-mu8}] for various decay modes of SM Higgs ($\gamma\gamma$ (yellow), $\tau \tau$ (cyan), $b\bar{b}$ (green), $ZZ^{\star}$ (red), $WW^{\star}$ (blue)) at 95$\%$ confidence level. The white shaded region simultaneously satisfies all the experimental constraints. 
 }
\label{4}
\end{figure}

First, we turn off all the dimension 6 operators (in Yukawa sector or EW gauge sector) and explore the effect of $g^{(6)}$ only. This is shown in upper left segment of Fig. \ref{4}. We find that to satisfy the constraints from Higgs observables, either $g^{(6)}$ has to be very small $\sim 0$ or new physics scale has to be very large. For an example, $g^{(6)}$ can be as large as 0.06 and as low as -0.06 for the new physics scale, M, to be 2 TeV. Since $g^{(6)}$ is responsible for both single Higgs and di-Higgs boson production simultaneously, dimension-6 term  ($g^{(6)}$) is highly constrained to give large di-Higgs production. For three of the benchmark points (BP1 and BP2), noted in upper left segment of Fig. \ref{4}, $\mu_{hh}$ and $\mu_{t\bar{t}h}$ is almost 1 and there is no significant deviation from SM prediction. Then, we add the contribution from dimension-6 Yukawa terms and we get a large region of the parameter space which is consistent with the Higgs observables and also gives significant deviation in $t\bar{t}h$ and $hh$ production. Due to the $g^{(6)}$ term, there will be two dominant processes for single Higgs production via ggF mode, one is due to the triangular loop circulated by top quark and the other one due to contact interaction term ($ggh$) and there will be large interference between these two diagrams.  Upper right segment of Fig. \ref{4} depicts the constraints in  $\lbrace y_{t}^{(6)}, g^{(6)}\rbrace $ plane from the signal strength limits [cf. Table \ref{table-mu8}] for various decay modes of SM Higgs at 95$\%$ confidence level. It is clear that when $y_{t}^{(6)}$ gets positive values, $g^{(6)}$ prefers negative values to compensate the overall enhancement effect in single Higgs production and vice versa. For two of the benchmark points (BP3 and BP4), the signal strength $\mu_{t\bar{t}h}$ becomes 2.0 and 0.7 and di-Higgs production cross-section becomes 64 fb and 41.8 fb respectively. Similarly, Lower left segment of Fig. \ref{4} depicts the constraints in  $\lbrace y_{b}^{(6)}, g^{(6)}\rbrace $ plane from the signal strength limits. Here we have fixed the value of $y_{t}^{(6)}$(=1) and new physics scale M (=500 GeV). In the survived parameter space, we choose three benchmark points (BP5 and BP6 as noted in this fig.) and calculate the $t\bar{t}h$ and $hh$ production rate. For benchmark points (BP5  and BP6), $\mu_{t\bar{t}h}$ equals 1.55 and di-Higgs production cross-sections are 31 fb and 81  fb respectively. Similarly, Lower Right segment of Fig. \ref{4} shows the constraints from the signal strength limits in  $\lbrace y_{g}^{(6)}, g^{(6)}\rbrace $ plane, where  we have kept a fixed value of , $y_{b}^{(6)}$(=-0.2) and new physics scale M (=500 GeV). For two of the benchmark points (BP7 and BP8), signal strengths ($\mu_{t\bar{t}h}$) become 1.3 and 1.6 and di-Higgs production cross-sections become 221 fb and 52 fb respectively. As we already mentioned, since dimension 6 term in the strong sector, responsible for di-Higgs production, is not decoupled from the term responsible for the single Higgs production, di-Higgs production rate can not be enormously large, but it can be  as large as 6 times of the SM predicted cross-section.

\begin{figure}[]
$$
  \includegraphics[height=6cm, width=0.5\textwidth]{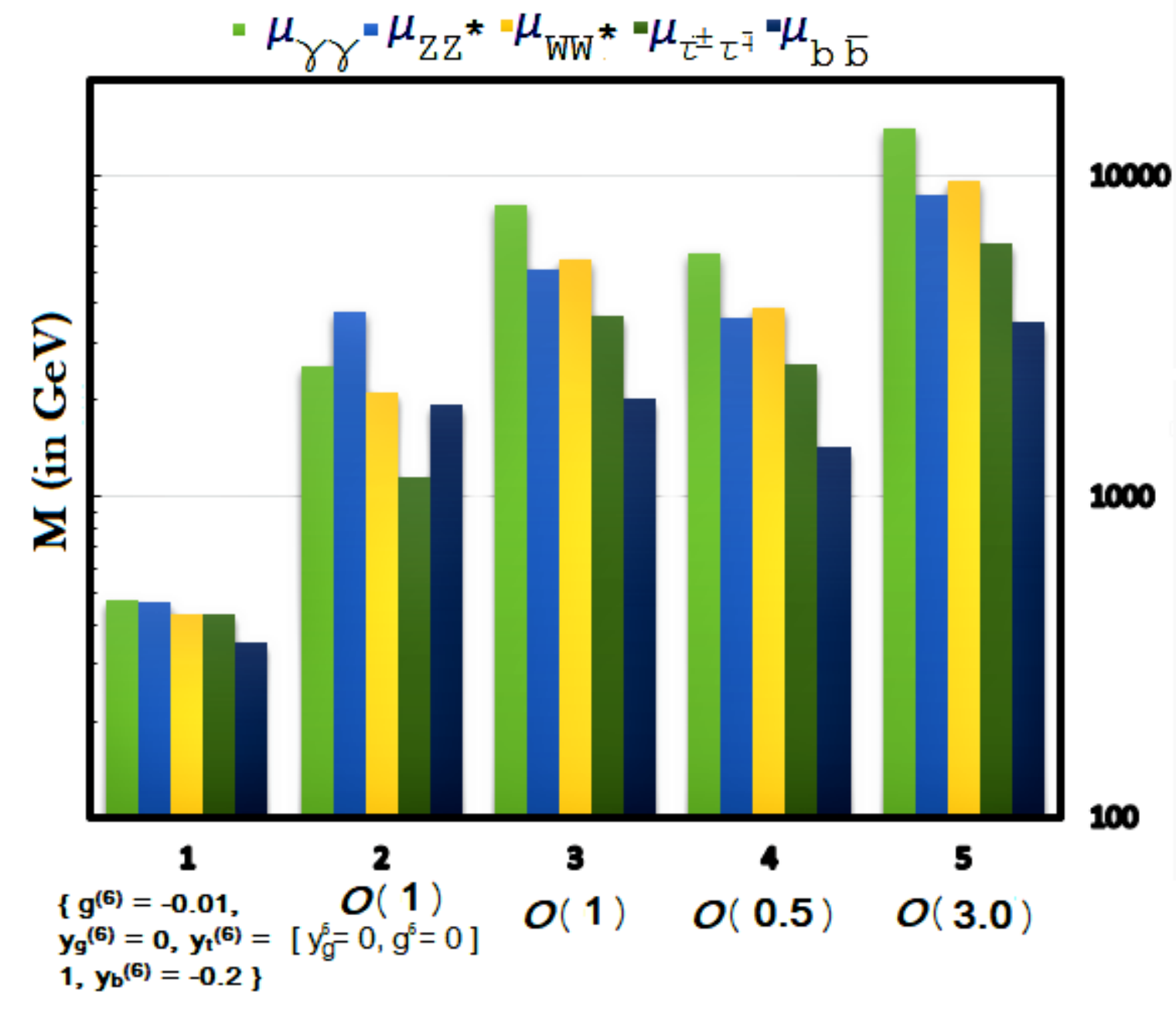}
 $$
 \caption{Estimation of new physics scale consistent with the measurements of Higgs observables.
}
\label{x}
\end{figure}

Now, we try to emphasize on the determination of the mass scale where these dimension-6 operators are generated and which is consistent with the measurement of Higgs observable.  We mention that the contribution of any effective operator is only sensitive to the ratio $g_{effective}/M^2$, and hence, new physics scale M is not observable without extra assumptions on the strength of the couplings $g_{effective}  \lbrace y_{t,b,g}^{(6)}, g^{(6)}, \lambda^{(6)}\rbrace$. In order to set limit on the new physics scale M, we have assumed $g_{effective}$ to be less than 3.5 to satisfy the perturbativity constraint. In Fig. \ref{x}, we have shown the limits on the mass scale M for different sets of the values of the effective six dimensional couplings. As we can see from Fig.\ref{x}, if all the dimension-6 couplings ($g_{effective}$) are $ \sim \mathcal{O}(3)$, new physics scale (M) up to 14 TeV is ruled out by the LHC Run II data [cf. Table \ref{table-mu8}] of the Higgs observables. Similarly, when all the dimension-6 couplings ($g_{effective}$) are $ \sim \mathcal{O}(1)$ [$ \sim \mathcal{O}(0.5)$], new physics scale (M) has to be at least 8 TeV [5.7 TeV] to be consistent with the LHC Higgs results [cf. Table \ref{table-mu8}] of Higgs searches. On the other hand, if we turn off dimension-6 term in strong sector and in EW gauge sector, the new physics scale can be much lower (3.7 TeV),  setting the dimension-6 term in yukawa sector $ \sim \mathcal{O}(1)$. Now we numerically scan the whole parameter space and we find that for a judicious choice of parameter space ($g^{(6)} = -0.01, y_{g}^{(6)} = 0, y_t^{(6)} = 1, y_b^{(6)} = -0.2$), the new physics scale M can be as low as 478 GeV satisfying all the Higgs physics constraints and giving new physics effect of enhanced $t\bar{t}h$ production. The reason is that there is a negative interference effect between two diagrams  contributing to the single Higgs production in ggF process (one is due to  effective $ggh$ coupling via triangular loop circulated by top quark and other one is the contact interaction term $ggh$ due to dimension-6 operator),  if $g^{(6)}$ has negative values. Hence, enhanced $t\bar{t}h$ coupling compensate that factor satisfying all the Higgs constraints and as a result, we can get enhanced $t\bar{t}h$ production. This scenario can be realized, if any colored particle contributes to the triangular loop in addition to the top and bottom quarks.  

We now clarify some points regarding the mass scale, and the limitation used for the six dimensional couplings.  Current LHC Higgs observables data are in agreement with the SM, so in principle all the 6-dimensional couplings can be zero. In that case, it is not possible to say anything about the scale of new physics. However, the Higgs observables still have large errors, and hence gives the possibility of the existence of new physics. The questions we have addressed is whether in this effective coupling parameter space, there are regions which are allowed by the data, and allow low scale of new physics as well as giving some new physics signatures such as enhanced $t\bar{t}h$ and $hh$ predictions. For example,  regarding the new physics scale, we want to mean that the new physics scale can be as low as 478 GeV making consistent with Higgs properties and also giving associated new physics signals like enhanced $t\bar{t}h$ or $hh$ predictions which can be testable at the current or upcoming run of LHC. Regarding the restriction on the effective couplings, since we consider the lowest order contributions, the higher order contributions will no longer be small, if the values of the couplings exceeds the perturbativity limit.  This gives a reasonable justification to our assumption.

\begin{figure}[]
$$
  \includegraphics[height=4.5cm, width=0.45\textwidth]{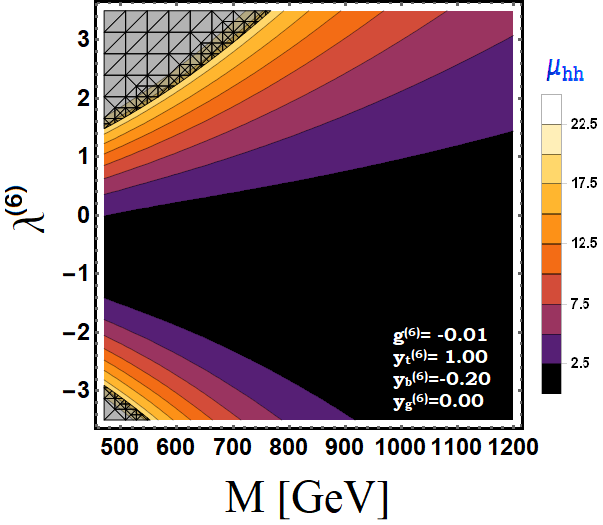}
  $$
  $$
  \includegraphics[height=4.5cm, width=0.45\textwidth]{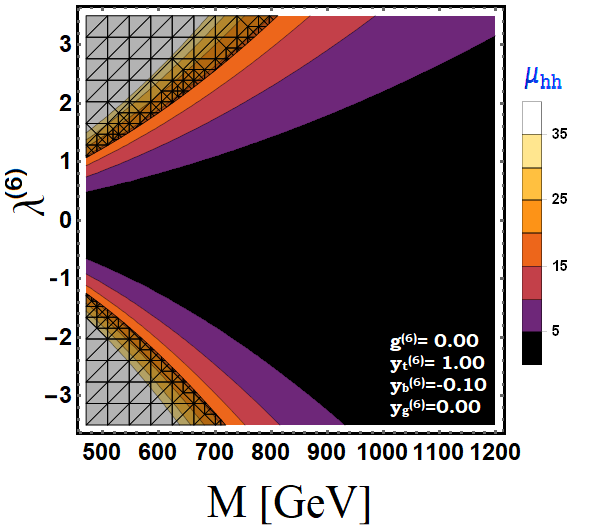}
 $$
 \caption{ Contour plot of signal strength of di-Higgs production $\mu_{hh}$ in $\lbrace \lambda^{(6)}, M\rbrace $ plane. Black meshed zone is excluded from current di-Higgs searches at the LHC. Scaling of $\mu_{hh}$ is shown on the right side of each figure. Left : $(g^{(6)} = -0.01, y_{g}^{(6)} = 0, y_t^{(6)} = 1, y_b^{(6)} = -0.2)$; Right : $(g^{(6)} = 0, y_{g}^{(6)} = 0, y_t^{(6)} = 1, y_b^{(6)} = -0.1)$. }
\label{y}
\end{figure}

Next,  we analyze the new physics contributions of the dimension-6 operator in the Higgs potential which contributes to the cubic Higgs coupling.
Due to the addition of the effective dimension six operator in the Higgs potential, the effective triple Higgs coupling is modified significantly as shown in Eq. 22. As a result, this has the most major effect on the di-Higgs production at the LHC. We take two set of benchmark points which allow enhanced $t\bar{t}h$ production rate (1.5 times of the SM predicted value) at the LHC making consistent with the Higgs properties. It is quite interesting to see from Fig. \ref{y} that we can get the signal strength $\mu_{hh}$ as big as 19 which means that the di-Higgs production cross-section can be as big as 636 fb which is 19 times of the SM predicted cross-section. We  mention that the di-higgs production cross-section can be even  larger than 636 fb for a certain region of parameter space as we can see from Fig. \ref{y}.  But, ATLAS and CMS collaborations have analyzed and reported the new results on di-Higgs boson searches \cite{econf, mor,CMShh,CMShh2,ATLAShh} looking at the different final states ($b\bar{b}\gamma\gamma, b\bar{b}\tau^{+} \tau^{-}, b\bar{b}b\bar{b}$ and  $b\bar{b}W^+W^-$), using 36 fb$^{-1}$ data from Run II of LHC at 13 TeV. Due to non-observation of any signal, the stringent limit of 636 fb on di-Higgs production cross section is reported \cite{econf, mor,CMShh,CMShh2,ATLAShh}. The black meshed zone in Fig. \ref{y} is excluded from this current di-Higgs searches. If LHC luminosity is upgraded to 3 $ab^{-1}$, SM like double Higgs production (33.45 fb) can be observed with $3.6 \sigma$ significance \cite{tg}. On the other hand, in our scenario, the enhanced di-Higgs production can be even sensitive to the 50 fb$^{-1}$ LHC luminosity which is close to the data set currently analyzed. We think this a very interesting scenario which simultaneously provides a testable smoking gun signal for the di-Higgs production and enhanced $t\bar{t}h$ production at the LHC. 
The future hadron-hadron circular collider (FCC-hh) or the super proton-proton collider (SppC), designed to operate at the energy of 100~TeV, can easily probe most of the parameter space in our scenario through the $hh$ pair production~\cite{ATL-PHYS-PUB-2014-019,Gomez-Ceballos:2013zzn,CEPC-SPPCStudyGroup:2015csa,tg}. As mentioned, the di-Higgs production in some sets of the six dimensional parameter space can be large enough to be observable even in this run of the LHC.

\section{Conclusion}\label{sec:con}
\vspace{-0.1cm}
In this work, we have made an investigation on the effect of the effective dimension six operators for the single Higgs productions, and the corresponding $\mu_{t\bar{t}h}$, as well as di-Higgs signals at the LHC.
Since the number of the effective dimension six operators are too many,
we have made a judicious choice of few operators which has the maximum impact for these observable. Using the experimental data at the LHC, we have analyzed in some detail the effects of these operators, how large or small the  $\mu_{t\bar{t}h}$, and di-Higgs signals can be, and how small the new physics scale can be satisfying all the available experimental constraints. We find the the $\mu_{t\bar{t}h}$ signal can be as large as two times of that in the SM, while the di-higgs production cross section can be as large as 19 times of that in the SM at the 13 TeV LHC with a new physics scale, M equal to $478$ GeV. These predictions can be tested as more data accumulates at the current and the future runs at the LHC. 
The results presented here can be taken as an initial guide in the exploration of the enhanced $t\bar{t}h$ and $hh$ signal at the LHC via dimension-6 operators. 
\vspace{-0.6cm}
\section*{Acknowledgement}
\vspace{-0.2cm}
We want to thank the anonymous referee for many important comments which  has helped us to improve the paper in a significant way. This work is supported in part by the US Department of Energy Grant No. de-sc0016013. The work of S.J. is also supported in part by the Fermilab Distinguished Scholars Program. SJ thanks the Fermilab Theoretical Physics Department for warm hospitality during the completion of this work.  SN
 thanks the warm hospitality of the Rice University HEP Group (where he is currently a visiting professor) during the completion of this work.



\end{document}